\documentclass{jpp}

\usepackage{graphicx}
\usepackage{amsmath}	% Advanced maths commands
\usepackage{amssymb}	% Extra maths symbols
\usepackage{multicol}        % Multi-column entries in tables
\usepackage{pdflscape}	% Landscape pages
\usepackage{gensymb}
\usepackage{tikz}
\usetikzlibrary{arrows}
\usepackage{multirow}

\newcommand{\kperp}{k_\perp}

\newcommand{\beq}{\begin{equation}}
\newcommand{\eeq}{\end{equation}}

\newcommand{\bega}{\begin{align}}
\newcommand{\eal}{\end{align}}
\newcommand{\ddt}{\frac{d}{dt}}
\newcommand{\toe}{T_{0{e}}}
\newcommand{\bdg}{\hat{\mathbf{b}}\cdot\nabla}
\newcommand{\me}{m_{e}}
\newcommand{\de}{d_{e}}
\newcommand{\nablap}{\nabla_\perp}
\newcommand{\apar}{A_\parallel}
\newcommand{\dtpare}{\delta T_{\parallel {e}}}
\newcommand{\gel}{g_{e}}
\newcommand{\vthe}{v_{\mathrm{th}{e}}}
\newcommand{\zot}{\frac{Z}{\tau}}
\newcommand{\din}{\delta_{\mathrm{in}}}

\newcommand{\toz}{\frac{\tau}{Z}}

\newcommand{\pxpx}{\partial_x^2}
\newcommand{\vay}{v_{\mathrm{A}y}}
\newcommand{\vaey}{v_{\mathrm{A}ey}}

\shorttitle{The onset of electron-only reconnection}

\shortauthor{A. Mallet}

\title{The onset of electron-only reconnection}
\author{Alfred Mallet\aff{1}
  \corresp{\email{alfred.mallet@berkeley.edu}}
}
\affiliation{\aff{1}Space Sciences Laboratory, University of California, Berkeley CA 94720, USA
}
\date{\today}

\begin{document}
\maketitle
\begin{abstract}
Motivated by recent observations of ``electron-only" magnetic reconnection, without an ion-scale sheet or ion outflows, in both the Earth's magnetosheath and in numerical simulations, we study the formation and reconnection of electron-scale current sheets at low plasma beta. 
We first show that ideal sheets collapse to thicknesses much smaller than the ion scales, by deriving an appropriate analogue of the Chapman-Kendall collapse solution. 
Second, we show that, in practice, reconnection onset happens in these collapsing sheets once they reach a critical aspect ratio, because the tearing instability then becomes faster than their collapse timescale. We show that this can happen for sheet thicknesses larger than the ion scale or at only a few times the electron scale, depending on plasma parameters and the aspect ratio of the collapsing structure, thereby unifying the usual picture of ion-coupled reconnection and the new regime of electron-only reconnection. %and thus the usual picture of ion-coupled reconnection and the new regime of electron-only reconnection can be unified. 
%Finally, we also show that reconnection onset happens at sheet thicknesses larger than the width of a steady-state reconnecting configuration, similar to recent results in the theory of resistive reconnection.
 We derive relationships between plasma beta, ion-to-electron temperature ratio, the aspect ratio, electron outflow velocity, and the final thickness of the sheets, and thus determine under what circumstances electron-scale sheets form and reconnect.
\end{abstract}

\section{Introduction}
Magnetic reconnection in current sheets is present in many astrophysical and laboratory settings and is an important process for converting magnetic energy into thermal and non-thermal kinetic energy \citep{zweibel2009,yamada2010,ji2019}. In low-$\beta$ collisionless guide-field reconnection, field lines are broken within a microscopic electron diffusion region, of size comparable to the electron inertial length $d_e = c/\omega_{pe}$, where $\omega_{pe}=\sqrt{4\pi n_e e^2/m_e}$ is the electron plasma frequency. In the usual picture of (guide-field) reconnection, this electron region is embedded in a larger-scale ``ion region" of thickness of order the ion sound scale $\rho_s= \rho_i\sqrt{ZT_e/2T_i}$, where the ion gyroradius $\rho_i= v_{{\rm th} i}/\Omega_i$, the ion thermal speed $v_{{\rm th} i} = \sqrt{2T_i/m_i}$ and the ion gyrofrequency $\Omega_i = ZeB_0/m_i c$. Associated with this region one expects bidirectional jets of plasma flowing away from the reconnection site at around the in-plane Alfv\'en speed $\vay = B_y/\sqrt{4\pi n_i m_i}$, where $B_y$ is the reconnecting (in-plane) magnetic field. We will term this standard picture ``ion-coupled" reconnection. 

However, recent analysis of data from the MMS spacecraft \citep{phan2018} has shown that, at least in Earth's turbulent magnetosheath, one overwhelmingly finds current sheets with thicknesses of only a few $d_e$ (i.e., much thinner than $\rho_s$) \emph{without} associated ion regions and ion jets, but with clear electron jets moving at appreciable fractions of the electron Alfv\'en velocity $\vaey= B_y/\sqrt{4\pi n_e m_e}$. This new regime has also recently been reproduced in the numerical simulations of \citet{pyakurel2019}. A similar situation has been observed in the hybrid Vlasov-Maxwell turbulence simulations of \citet{califano2018}, who observed that when energy is injected at scales much larger than the ion scales, ion-coupled reconnection is observed, while if energy is injected closer to the ion scales, only electron jets and electron-scale sheets are observed. Reconnection of current sheets with thicknesses of a few $d_e$ is also observed in the VINETA-II experiment \citep{jain2017}. We will follow the terminology of \citet{califano2018} and term this new regime ``electron-only" reconnection.

The goal of this paper is to develop a theoretical picture that can explain the onset of this new regime of electron-only reconnection, and under what circumstances it occurs in place of the standard ion-coupled reconnection. To achieve this, we will make use of a simplified model, rigorously derived from gyrokinetics in the limit $\beta_e \sim m_e/m_i\ll 1$, the so-called ``kinetic reduced electron heating model", henceforth KREHM \citep{zocco2011}. The details of this model, as well as its utility and limitations, are described in Section \ref{sec:eq}. 

To make progress in understanding electron-only reconnection onset, there are two main questions which should be answered. First, can current sheets collapse to thicknesses of order a few $d_e$, significantly below the ion scales? We answer this in Section \ref{sec:collapse} by deriving a collapsing X-point solution analogous to Chapman-Kendall collapse \citep{chapman1963,biskampbook}, and showing that the solution is valid both above and below the ion scales. This means that ideal (i.e. with field lines frozen into an effective electron flow) sheets can indeed collapse well past the ion scale.

Second, at what sheet thickness do these collapsing sheets begin to reconnect, and under what conditions is this thickness above (obtaining ion-coupled reconnection) or well below (obtaining electron-only reconnection) the ion scales? We answer this in Section \ref{sec:disrupt}, showing that the sheets are disrupted when they collapse to a critical thickness\footnote{Notably, this critical thickness at which reconnection onset occurs is always larger than that of the steady-state reconnecting configuration, derived in Appendix \ref{sec:sp}, similar to the results of \citet{pucci2014} and \citet{uzdensky2016} for resistive reconnection onset.} at which the linear tearing growth rate\footnote{We derive the growth rate of the tearing mode for sheets with thicknesses below the ion scales in Appendix \ref{sec:tear}.} is faster than the collapse or formation rate of the sheet derived in Section \ref{sec:collapse}. Our analysis allows us to derive relationships between $\beta_e$ and the aspect ratio and thickness of the sheet at disruption (reconnection onset), and we thus determine the physical conditions under which one will observe ion-coupled or electron-only reconnection. The essential physical result is that if the initial collapsing solution is not sufficiently anisotropic when its thickness is above the ion scales, the collapse continues down to a thickness well below the ion scale, and electron-only reconnection will occur. Remarkably (considering the formal limitation of KREHM to low $\beta$), we will see that our model seems to agree rather well with the magnetosheath observations of electron-only reconnection \citep{phan2018,stawarz2019}.

%In this paper, we develop a theoretical picture for how and under what conditions these electron-scale current sheets form, and under what conditions they might reconnect. We begin by outlining the equations we will use for our analysis in Section \ref{sec:eq}. In Section \ref{sec:sp}, we show an equivalent of the Sweet-Parker \citep{sweet1958,parker1957} solution exists for sheets of thickness $d_e$ in our system. In Section \ref{sec:collapse}, we show that current distributions can collapse to sheets much thinner than the characteristic ion scales, with a solution analogous to the Chapman-Kendall X-point collapse \citep{chapman1963} for incompressible MHD. In Section \ref{sec:disrupt}, we argue that this collapse stops before the solution of Section \ref{sec:sp} is reached, because forming sheets are disrupted by the tearing instability and the subsequent onset of reconnection \citep{uzdensky2016} before they thin to $d_e$. We also derive relationships between $\beta_e$ and the aspect ratio and thickness of the sheet at disruption, and thus determine the physical conditions under which one might expect to observe ion-coupled and electron-only reconnection.
\section{Equations}\label{sec:eq}
Our starting point is the ``kinetic reduced electron heating model" (KREHM) derived by \citet{zocco2011} (henceforth ZS11) from gyrokinetics for electron beta of order the mass ratio $\beta_e \sim Z m_e/m_i$, where $Z=q_i/e$ (and the temperature ratio $\tau=T_{0i}/T_{0e}\sim 1$). The spatial scales of the fluctuations are ordered to be comparable  to the relevant ion kinetic scales, $k_\perp\rho_i \sim k_\perp\rho_s \sim1$, and so KREHM is able to capture both the large-scale ($k_\perp\rho_s\ll1$) and small-scale ($k_\perp\rho_s\gg1$) behaviour of the plasma, as well as the transition between them. This is essential for our present purposes, since we are attempting to diagnose the transition between ion-coupled and electron-only reconnection onset.

KREHM is in some sense a minimal model for describing collisionless guide-field reconnection, since the equations incorporate the dispersive effects entering at the ion sound scale $\rho_s$ that are thought to be essential to enable fast reconnection \citep{rogers2001}, a flux-unfreezing mechanism (electron inertia) entering at the electron inertial scale $\de$, and a rigorous treatment of the (electron) heating channel. The equations, written in terms of the electrostatic potential $\phi$, the magnetic potential $\apar$, and the reduced electron distribution function $g_e$, are
\begin{align}
\ddt \frac{Z}{\tau}\left(1-\hat{\Gamma}_0\right) \frac{e\phi}{\toe} &= \bdg\frac{e}{c\me} \de^2 \nablap^2 \apar,\label{eq:mom}\\
\ddt \left( \apar - \de^2 \nablap^2\apar \right) &= - c\frac{\partial \phi}{\partial z} - \frac{c \toe}{e}\bdg \left[ \frac{Z}{\tau}\left(1-\hat{\Gamma}_0\right)\frac{e\phi}{\toe} - \frac{\dtpare}{\toe}\right],\label{eq:apar} \\
\frac{d \gel}{dt} + v_\parallel \bdg \left(\gel - \frac{\dtpare}{\toe}F_{0\mathrm{e}} \right) &= C[\gel] + \left(1-\frac{2 v_\parallel^2}{\vthe^2}\right)F_{0\mathrm{e}} \bdg\frac{e}{c\me} \de^2 \nablap^2 \apar,\label{eq:ge}
\end{align}
where
\begin{align}
\ddt &= \frac{\partial}{\partial t} + \frac{c}{B_0} \left\{ \phi, ...\right\},\\
\bdg &= \frac{\partial}{\partial z} - \frac{1}{B_0} \left\{ \apar, ...\right\},\\
\frac{\dtpare}{\toe} &= \frac{1}{n_{0\mathrm{e}}}\int d^3\mathbf{v} \frac{2 v_\parallel^2}{\vthe^2} \gel,
\end{align}
and $\hat{\Gamma}_0$, encoding the ion FLR effects, is the inverse Fourier transform of
\beq
\Gamma_0\left(\frac{1}{2}\kperp^2\rho_i^2\right)=I_0\left(\frac{1}{2}\kperp^2\rho_i^2\right) e^{-\frac{1}{2}\kperp^2\rho_i^2},
\eeq
where $I_0$ is the modified Bessel function. The electron density perturbation $\delta n_e$ is related to $\phi$ via the relation
\beq
\frac{\delta n_e}{n_{0e}} = -\zot\left(1-\hat{\Gamma}_0\right)\frac{e\phi}{T_{0e}}.
\eeq
The Poisson bracket appearing in the equations is $\{f,g\} = \hat{\mathbf{z}}\cdot \nabla_\perp f \times \nabla_\perp g$. We have taken the limit $\nu_{ei} \ll \omega$, which allows us to neglect a resistive term in Eq.~(\ref{eq:apar}) (but not the collisions $C[\gel]$ in Eq.~\ref{eq:ge}). 

The function $g_e$ is a mathematically convenient reduced electron-gyrocenter parallel velocity distribution function, which encodes the electron heating channel. This is perhaps all the information that is needed to understand the present paper: in more detail, it is related to the perturbed distribution function $\delta f_e$ by
\beq
\delta f_e(\mathbf{r},\mathbf{v},t) = \left(1 + \frac{\delta n_e}{n_{0e}} + \frac{2 v_\parallel u_{\parallel e}}{\vthe^2} \right)F_{0e} + g_e(\mathbf{R}_e,\mathbf{v},t)
\eeq
where $\mathbf{R}_e=\mathbf{r}+\frac{\mathbf{v_\perp}\times \hat{\mathbf{z}}}{\Omega_e}$ is the electron gyrocenter. Integrating $0$th and $1$st $v_\parallel$-moments of $\delta f_e$ over all velocity (perturbed electron density and parallel flow), we can see that contribution from $g_e$ to these vanishes, i.e. $g_e$ contains the information of all higher moments, but not density or parallel flow. It turns out (see ZS11 for details) that the non-trivial $v_\perp$ dependence of $g_e$ can be safely ignored, and so $g_e$ can be considered a function of $v_\parallel$, space and time only.

We may write Eq.~(\ref{eq:apar}) as
\beq
\frac{\partial\apar}{\partial t} = -c\bdg\left[\phi + \zot\left(1-\hat\Gamma_0\right)\phi -\frac{\dtpare}{e}\right]+\frac{d}{dt}\de^2\nablap^2\apar\label{eq:apar2},
\eeq
which defines an effective velocity 
\beq
\mathbf{u}_{\rm eff} = \hat{z}\times \nablap \frac{c}{B_0}\left[\phi + \zot\left(1-\hat\Gamma_0\right)\phi -\frac{\dtpare}{e}\right],\label{eq:frozen}
\eeq
into which the field lines are frozen, apart from the effects of electron inertia \citep{zocco2011}. Thus, field lines are only broken when gradients are of order $d_e^{-1}$.

These equations have various interesting limits. First, for fluctuations on long lengthscales(compared to the ion gyroradius) $k_\perp\rho_i \ll 1$, and $\Gamma_0 \approx 1-k_\perp^2\rho_i^2/2$, and one can show that the equations reduce to those of RMHD, reduced magnetohydrodynamics \citep{strauss1976}, describing nonlinearly interacting Alfv\'enic fluctuations propagating up and down the magnetic field. Second, if one takes $k_\perp\rho_i \gg1$, $\Gamma_0 \approx 0$, and subsequently expand in the square root of the mass ratio $\sqrt{m_e/m_i}$, i.e. $k_\perp d_e \ll1$, one obtains the equations of ERMHD, electron reduced magnetohydrodynamics \citep{schektome2009}, which involve nonlinearly interacting kinetic-Alfv\'en fluctuations propagating up and down the field. %Finally, in this paper we will neglect the parallel spatial dimension, i.e. assume that $\partial/\partial z \ll \partial/\partial x, \partial/\partial y$.

It is important to note that, as mentioned earlier, formally these equations are only valid for $\beta_i \sim \beta_e \ll 1$. This rather limits their formal applicability to some physical regimes, including the magnetosheath, where electron-only reconnection has recently been observed \citep{phan2018,stawarz2019}, where $\beta_i \gtrsim1$ and $\beta_e \lesssim1$. Nonetheless, we believe that their simplicity allows us to model the essential physics of current sheet collapse and reconnection onset in a useful way, for three main reasons. First, simulations of turbulence using KREHM have revealed that it seems to compare rather well with more complete models, even at higher $\beta$ than its formal regime of applicability \citep{groselj2017}.
Second, we will show in Section \ref{sec:collapse} that the Chapman-Kendall-like current sheet collapse solution also exists in ERMHD at arbitrary $\beta_i$ and $\tau$, meaning that this part of the analysis remains true even without the formal $\beta_i\ll 1$ requirement. 
Finally, the collisionless tearing mode (which we will show enables the onset of reconnection in Section \ref{sec:disrupt}) has a growth rate that decreases with $\beta_e$ \citep{numata2015} for constant ratio of sheet thickness to $\rho_i$, meaning that at higher $\beta_e$, the basic physics of our picture will remain, but with the sheets potentially attaining even smaller (relative to $\rho_i$) thicknesses.

We should also mention here that \citet{boldyrevloureiro2019} have used a set of fluid equations derived for $k_\perp\rho_i\gg1$, $\beta_i\sim1$, $\beta_e\ll1$ to study the small-scale tearing instability and its effect on turbulence, obtaining some results similar to ours (which is an additional source of reassurance that our low-$\beta_i$ ordering does not invalidate the basic picture of tearing-induced reconnection onset): we have used KREHM instead because, first, we want a set of equations covering the ion-scale transition at $k_\perp\rho_i \sim 1$ to describe the transition between the ion-coupled and electron-only reconnection regimes, and second, the electron heating channel as encoded by Eq.~(\ref{eq:ge}) is important for a realistic picture of reconnection.

\section{X-point collapse}\label{sec:collapse}
For electron-only reconnection to occur, it must first be possible to form current sheets with thicknesses well below the ion scale. In ideal MHD, it is well known that X-points tend to collapse into sheets \citep{syrovatskii1981}. It turns out that this is still true in our system of equations, even at scales well below the ion gyroradius. If we require 
\beq
\bdg\nablap^2\apar=0,\label{eq:equil}
\eeq
then $\gel=0$ is a solution. Let us also impose a ``vorticity-free" condition on the system, 
\beq
\nablap^2\phi=0.\label{eq:vortfree}
\eeq
Under this condition, $\hat\Gamma_0\phi = \phi$\footnote{The series defining the modified Bessel function is $I_0(x) = \sum_{m=0}^{\infty} \frac{(x/2)^{2m}}{(m!)^2}$, so $\hat\Gamma_0$ is just a series in $\nablap^2$, with the $m=0$ term being $1$. If we had first taken the small-scale limit $k_\perp\rho_i\gg1$ and said that $\hat\Gamma_0\approx 0$ using the large-argument asymptotic expansion of $I_0$, we would have missed this fact.}, and along with (\ref{eq:equil}), the whole of Eq.~(\ref{eq:mom}) vanishes. Our flow must, of course, be consistent with maintaining the condition (\ref{eq:equil}). 

%If we take the $\kperp\de \ll1$ limit of Eqs.~(\ref{eq:mom}--\ref{eq:ge})\footnote{The range of scales between $\rho_i$ and $d_e$ is rather limited, so this expansion may be questioned to some extent - however, it will allow us to understand physically the processes at work in the plasma}, the field lines are then perfectly frozen into $\mathbf{u}_{\rm eff}$ (cf. Eq.~\ref{eq:frozen}), and o
\subsection{Chapman-Kendall solution}
One (2-dimensional) solution which satisfies the conditions (\ref{eq:equil},\ref{eq:vortfree}) is a slightly modified version of ``Chapman-Kendall" collapse \citep{chapman1963,biskampbook}, originally derived for MHD (but which turns out to still be applicable here),
\begin{align}
\phi &= \frac{B_0}{c}\frac{\Lambda(t)}{2} x y,\label{eq:phick}\\
\apar &= \frac{B_{\perp}}{2}\left(\frac{x^2+2\de^2}{a(t)} - \frac{y^2+2\de^2}{L(t)}\right).\label{eq:aparck}
\end{align}
where $a(t)$ and $L(t)$ are the thickness and length of the current sheet respectively. This solution is an approximation valid in the region close to the X-point in the interior of a realistic current sheet. Inserting this into Eq.~(\ref{eq:apar2}), we obtain
\begin{align}
\dot a &= -\Lambda(t)a, \label{eq:a}\\
\dot L &= \Lambda(t)L.\label{eq:L}
\end{align}
Thus, in the absence of reconnection, current sheets can thin to scales much smaller than the ion scales. However, this does not show that such thin sheets are always physically realisable: we will show in Section \ref{sec:disrupt} that, in fact, localised current configurations with $L> a$ are disrupted by the onset of reconnection at scales larger than $d_e$ (but, importantly, sometimes smaller than $\rho_s$)\footnote{Indeed, Eqs.~(\ref{eq:phick},\ref{eq:aparck}) are valid even at scales below $d_e$: perhaps a hint that something interesting might happen at a larger scale which invalidates this simple scenario.}. \citet{white2018} have found that KREHM also allows a more general class of Chapman-Kendall-like solutions, for example a helically twisted version of the simple solution presented here.

\subsection{Choosing $\Lambda(t)$}
It remains for us to decide a reasonable form for $\Lambda(t)$. We do this in two limiting cases, $a\gg\rho_i$ and $a\ll\rho_i$, and ask that the outflows match the characteristic wave speed involved in maintaining the structure of the sheet in the $y$-direction in both cases. In the case where $a\gg\rho_i$, the outflows should attain the constant in-plane Alfv\'en speed $\vay = B_{y}(a)/\sqrt{4\pi n_{0i}m_i}$ at $y\sim L(t)$. In the small-scale case where $a\ll\rho_i$, the outflow speed at $y\sim L(t)$ should be the in-plane kinetic Alfv\'en wave speed (see ZS11 or \citealt{schektome2009}):
\beq
u_{\mathrm{KAW}y} = \sqrt{\frac{1}{2}\left(1+\zot\right)}\vay\frac{\rho_i}{a(t)}
\eeq
Importantly, this is inversely proportional to $a$%\footnote{%This outflow speed also matches that of the steady-state solution derived in Appendix \ref{sec:sp}, Eq.~(\ref{eq:uout}): although we will find (see Section \ref{sec:disrupt}) that the steady-state width is unlikely to be attained. 
%If the reader is skeptical about our appeal to causality, one may instead use the analysis of the first three equations of Appendix \ref{sec:sp} and show that this outflow speed (which also appears in Eq.~\ref{eq:uout}) is required simply by conservation of energy upstream and downstream of the X-point.}
. Thus, it is physically reasonable to choose
\beq
\Lambda(t) =\left\{ \begin{array}{cr}
\frac{2\vay}{L(t)}, & a \gg \rho_i,\\
\frac{\sqrt{2(1+Z/\tau)}\vay\rho_i}{a(t)L(t)}, &a\ll\rho_i,\\
\end{array}\right.\label{eq:lambda}
\eeq
which defines (inverse) timescales associated with forming sheets of thickness $a$ and length $L$. With these choices, Eqs. (\ref{eq:a}) and (\ref{eq:L}) have the solutions
\begin{align}
a\gg\rho_i:& %\quad a=a_0L_0/(L_0+2\vay t), 
\quad L=L_0+2\vay t,\quad a=a_0L_0/L,\label{eq:al_big}\\
a\ll\rho_i:& %\quad a=a_0 \exp\left(-\frac{\sqrt{2(1+Z/\tau)}\vay\rho_i}{a_0L_0}t\right), 
\quad L=L_0\exp\left(\frac{\sqrt{2(1+Z/\tau)}\vay\rho_i}{a_0L_0}t\right), \quad a=a_0L_0/L.\label{eq:al_small}
\end{align}
The aspect ratio as a function of $a$ is given by
\beq
\frac{L}{a} = \frac{a_0L_0}{a^2},\label{eq:aspscaling}
\eeq
simply from continuity. It is important to note that if the sheets are formed as part of turbulence, this scaling is not accurate, because structures will interact with each other as well as undergoing their own nonlinear evolution. Instead, one might expect, for $a\gg\rho_i$, $L/a \propto a^{-1/4}$ (cf. \citealt{boldyrev,chandran14,ms16}). The aspect ratio scaling for $a\ll\rho_i$ in turbulence is as yet unknown.

\subsection{Collapse time}
Let us start with a structure of thickness $a_0>\rho_i$, and ask how long it takes for it to collapse to a thickness $a_* < \rho_i$. First, using Eq.~(\ref{eq:al_big}), the time taken to collapse to $\rho_i$ is
\beq
t_\rho \sim \left(\frac{a_0}{\rho_i}-1\right)\frac{L_0}{2\vay} \approx \frac{a_0L_0}{2\vay\rho_i}\label{eq:trho}
\eeq
Subsequent collapse from $\rho_i$ down to $a_*$ occurs according to Eq.~(\ref{eq:al_small}) in a time
\beq
t_* \sim \frac{a_0L_0 \ln(\rho_i/a_*)}{\sqrt{2\left(1+\zot\right)}\vay\rho_i},\label{eq:tstar}
\eeq
which is, in practice, usually of order $t_\rho$.
Thus, the total collapse time is of order $t_\rho$, and mainly depends on the initial size and reconnecting field within the structure.

\subsection{Current sheet collapse at higher $\beta$}\label{sec:ermhdcollapse}
Finally, the question of whether or not this collapse process depends crucially on the low-$\beta$ ordering used to derive KREHM may be of interest to the reader. First, for $k_\perp\rho_i \ll 1$, the Chapman-Kendall collapse obviously works, since the RMHD equations are valid at arbitrary $\beta$ and can be derived from gyrokinetics under the assumption $k_\perp\rho_i\ll1$ \citep{schektome2009}. Second, for $k_\perp\rho_i\gg1$, we can examine the (2D) electron reduced MHD equations \citep{schektome2009}, which may be manipulated into the following form, chosen for their similarity to Eqs.~(\ref{eq:mom}) and (\ref{eq:apar}),
\begin{align}
\ddt \frac{Z}{\tau} \frac{e\phi}{\toe} &= \frac{2}{2+\beta_i\left(1+\zot\right)}\bdg\frac{e}{c\me} \de^2 \nablap^2 \apar,\label{eq:mom_fluid}\\
\ddt \apar &= -\frac{c \toe}{e}\bdg \left[ \frac{Z}{\tau}\frac{e\phi}{\toe}\right],\label{eq:apar_fluid}
\end{align}
The effective velocity into which the field lines are frozen is
\beq
\mathbf{u}_{\rm eff} = \hat{z}\times \nablap \frac{c}{B_0}\left(1+\zot\right)\phi,
\eeq
which can be compared with Eq.~(\ref{eq:frozen}).
Under the condition (\ref{eq:equil}), Eq.~(\ref{eq:mom_fluid}) vanishes\footnote{Note that the $d\phi/dt$ in Eq.~(\ref{eq:mom_fluid}) could in fact just be written $\partial\phi/\partial t$, because $\left\{\phi,\phi\right\}=0$.}. Inserting Eqs.~(\ref{eq:phick}) and (\ref{eq:aparck}) with $d_e=0$ into Eq.~(\ref{eq:apar_fluid}), and requiring the outflow velocity to be the in-plane kinetic-Alfv\'en velocity (at larger $\beta_i$, see \citealt{schektome2009}), %and using the $a\ll \rho_i$ expression for $\Lambda(t)$ in Eq.~(\ref{eq:lambda}), 
\beq
u_{\mathrm{KAW}y} = \sqrt{\frac{1+Z/\tau}{2+\beta_i(1+Z/\tau)}}\vay\frac{\rho_i}{a(t)},
\eeq
we find that the solutions for $a(t)$ and $L(t)$ are 
\beq
L=L_0\exp\left(\frac{\sqrt{\frac{2(1+Z/\tau)}{1+\beta_i(1+Z/\tau)/2}}\vay\rho_i}{a_0L_0}t\right), \quad a=a_0L_0/L,
\eeq
only mildly different to the KREHM case.
Thus, the fact that current sheets collapse even at thicknesses $a\ll\rho_i$ does not depend on the $\beta\ll1$ ordering of KREHM.
\begin{table}
\begin{center}
\begin{tabular}{lcccccc}
 $a/\rho_s$&$\Delta'\din$&$\gamma/k_y\vay$& $\delta$ & $\din$&$\gamma_{\rm max}a/\vay$&$k_{\rm max}a$  \\
 \hline
\multirow{2}{*}{$\gg 1$}&$\ll1$&$H\Delta' {\de\rho_s}/{a}$& $H\de^2\Delta'$&$H^{1/2}\de\rho_s^{1/2}\Delta'^{1/2}$& \multirow{2}{*}{$\frac{H\de^{1/3+2/3n}\rho_s^{2/3+1/3n}}{a^{1+1/n}}$}&\multirow{2}{*}{$\left(\frac{\de^{2/3}\rho_s^{1/3}}{a}\right)^{1/n}$}   \\
&$\gg1$&${H\rho_s^{2/3}\de^{1/3}}/{a}$&${H\de^{4/3}}{\rho_s^{-1/3}}$&$H^{1/2}\rho_s^{1/3}\de^{2/3}$&   \\
\multirow{2}{*}{$\ll 1$}&  $\ll1$&$H\Delta' {\de\rho_s}/{a}$& $H\de^2\Delta'$&$\delta$&\multirow{2}{*}{$\frac{H\rho_s}{a}\left(\frac{\de}{a}\right)^{1/n}$}&\multirow{2}{*}{$\left(\frac{\de}{a}\right)^{1/n}$}    \\
&($\gg1$&${H\rho_s}/{a}$&$H\de$&$\delta$)&    \\

\end{tabular}
\caption{Scalings for the collisionless tearing mode in various limits, as discussed in Appendix~\ref{sec:tear}. The final row, $a/\rho_s\ll1$ and $\Delta'\din \gg 1$, is probably inaccurate since we do not have an analytic solution. The quantity $H=\sqrt{1+\tau/Z}$; please see the Appendix for details.\label{tab:scalings}}
\end{center}
\end{table}
\section{Disruption of forming sheets by the tearing instability}\label{sec:disrupt}
A useful model of the disruption of forming current sheets by reconnection has been developed recently by \citet{uzdensky2016}. In this model, collapsing sheets are disrupted if their linear tearing mode is faster than their formation timescale, i.e.
\beq
\gamma/\Lambda(t) > 1,\label{eq:gammalambda}
\eeq
assuming that the nonlinear stage of the tearing mode is at least as fast as the linear stage. 

In general, the tearing mode that disrupts the sheet has the maximum tearing growth rate $\gamma_{\rm max}$ if it fits into the sheet, i.e. if $k_{\rm max}L>1$, and by the growth-rate of the longest-wavelength mode that fits in the sheet otherwise, i.e. the mode with $k_y L = 1$. The relevant tearing growth rates for our system are summarized in Table \ref{tab:scalings}: the growth rates for the case $a\gg\rho_s$ are derived in ZS11, while we derive the growth rates for the opposite case with $a\ll\rho_s$ in Appendix~\ref{sec:tear}. 
\subsection{Critical aspect ratio}
One can show that for our choice of $\Lambda(t)$, in the case $a\gg\rho_s$, $k_{\rm max}$ becomes accessible before the mode with $k_yL=1$ goes unstable, while in the case where $a\ll\rho_s$, the sheet is disrupted when $k_{\rm max} L \sim 1$, and thus in practice we may always use $\gamma_{\rm max}$. Using Eq.~(\ref{eq:gammalambda}), we find that sheets are disrupted for aspect ratios 
\begin{align}
\frac{L}{a} > \left(\frac{L}{a}\right)_{crit}\sim\left\{\begin{array}{cr}
%\frac{a^{1+1/n}}{\de^{1/3+2/3n}\rho_s^{2/3+1/3n}}
\frac{1}{\sqrt{1+\tau/Z}}\left(\frac{a}{\de}\right)^{1/n}\left(\frac{a}{\rho_s}\right)\left(\frac{\rho_s}{\de}\right)^{(1-1/n)/3}, & a\gg\rho_s\\
\left(\frac{a}{\de}\right)^{1/n},&a\ll\rho_s.\\
\end{array}\right.\label{eq:lacrit}
\end{align}
The $a\gg\rho_s$ expression was previously found by \citet{delsarto2016}\footnote{The expressions do not match for $a=\rho_s$: this is because the analysis of previous sections fails around $a\sim\rho_s$, where the operator $\hat\Gamma_0$ cannot be expanded in small or large argument.}.

%\footnote{In principle, this could be ``fixed" by using the exact in-plane kinetic-Alfv\'en velocity (cf. \citealt{pyakurel2019}) to define $\Lambda$, and the exact tearing dispersion relation: we forgo numerical analysis here and note that there must be a smooth transition between our two limiting regimes.}%For $n=1$, they do match: this is because for $n=1$, the small-$\Delta'$ growth rate is independent of $k_y$ and the same for both $a\gg\rho_s$ and $a\ll\rho_s$, so that $\gamma_{\rm max}$ is the same for all $a$ -- despite $k_{\rm max}$ being different.
\begin{figure}
\includegraphics[width=\linewidth]{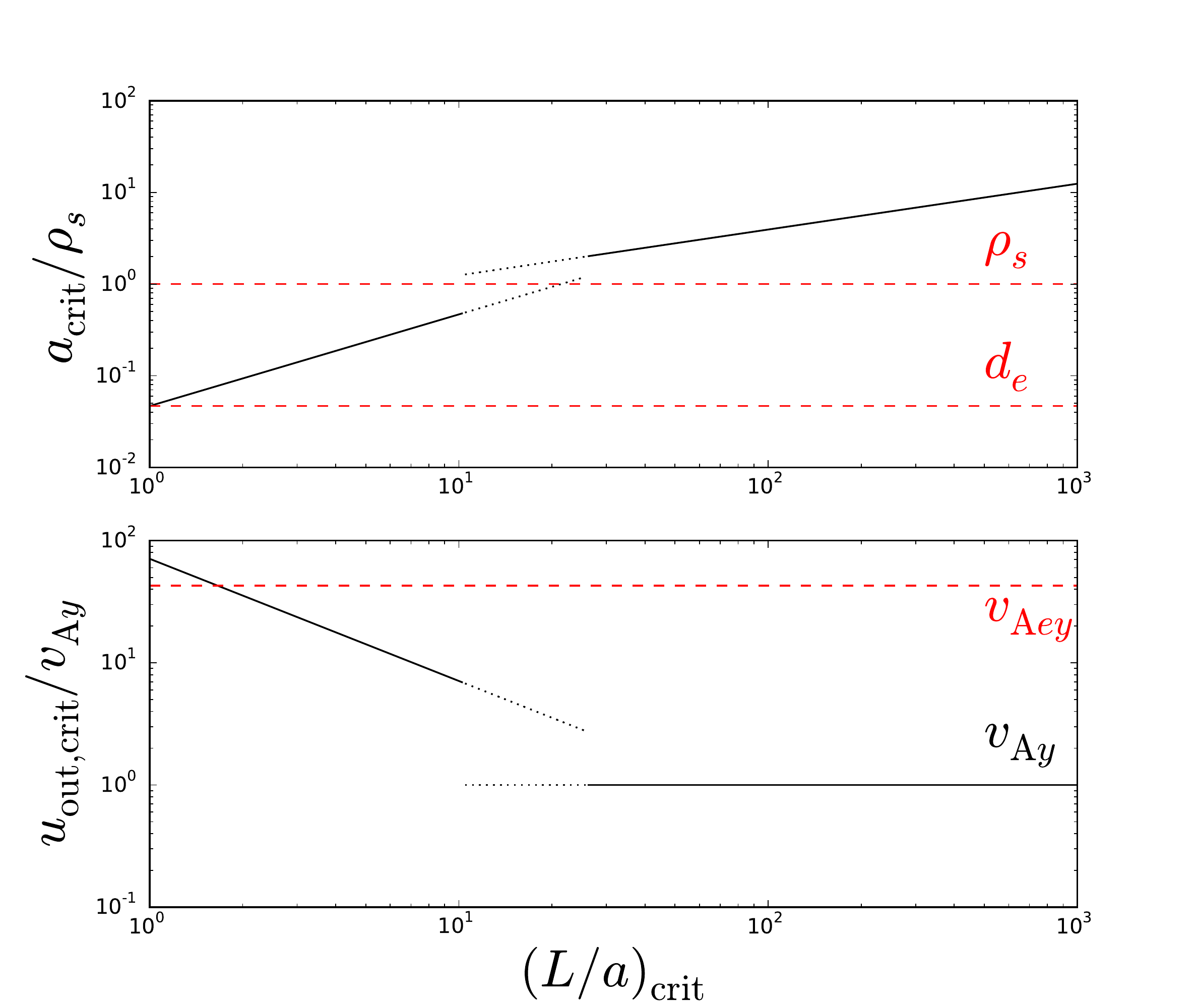}
\caption{In black, the disrupted thicknesses (top panel) and outflow velocities (bottom panel) of reconnecting sheets as a function of aspect ratio (cf. Eqs.~\ref{eq:lacrit} and \ref{eq:uout}). Because we have only derived scalings in the asymptotic limits $a\gg\rho_s$ and $a\ll\rho_s$, for $0.5<a_{\rm crit}/\rho_s<2$ we have plotted these two scalings as dotted lines; the true scaling must lie between them. Red dashed lines show the positions of $\rho_s$ and $d_e$ (top panel) and $\vaey$ (bottom panel). We have used parameters taken from \citet{stawarz2019}, $\beta_e=0.5$, $\tau=10$.\label{fig:disrupt}}
\end{figure}
%After the initial disruption of the sheet, there are $N\sim k_{\rm max}L$ new X-points, which can then subsequently collapse into new sheets with $L_{\rm new}\sim 1/k_{\rm max}$. For $a\gg\rho_s$, because $N\gg1$,
%\beq
%\frac{L_{\rm new}}{a} \sim 
%\left(\frac{a}{\de^{2/3}\rho_s^{1/3}}\right)^{1/n} \sim \frac{\rho_s}{a}\left(\frac{\de}{\rho_s}\right)^{1/3}\left(\frac{L}{a}\right)_{crit}\ll\left(\frac{L}{a}\right)_{crit},\quad a\gg\rho_s,
%\eeq
%they are initially stable. For $a\ll\rho_s$, because $N=1$, the new sheets are only marginally stable.

We have shown that there is a maximum achievable aspect ratio for sheets with both $a\gg\rho_s$ and $a\ll\rho_s$. Inserting $a\sim\rho_s$ into Eq.~(\ref{eq:lacrit}), sheets will become unstable before they reach the ion scales ($a\sim\rho_s$) only if their aspect ratio becomes larger than $(\rho_s/\de)^\nu$, where $1/n\leq\nu\leq1/3+2/3n$. Once the sheet reaches the thickness at which it disrupts, it reconnects at that thickness\footnote{As in the picture of ``ideal tearing" proposed by \citet{pucci2014} and elaborated in, e.g., \citet{tenerani2016} for resistive reconnection.} until all the available flux has been used up. Thus, if one observes a reconnecting (i.e. disrupting) sheet with length
\beq
L > \frac{\rho_s}{\sqrt{1+\tau/Z}} \left(\frac{m_e}{m_i}\frac{Z}{\beta_e}\right)^{\nu/2}, %\approx 10\rho_s,
\eeq
one expects to see that the sheet thickness is larger than the ion scale and that there should be bidirectional ion jets moving at $\vay$. This roughly agrees with the numerical simulations of \citet{pyakurel2019}, who found that to obtain fully ion-coupled guide-field reconnection, $L$ needed to be $10$s of $d_i$, and is reminiscent of the results of \citet{mandt1994} for reconnection without a guide field, who observed a change in the scaling of the reconnection rate once the length of sheets became comparable to a few times the ion inertial length. In the magnetosheath interval studied by \citet{phan2018} and subsequently \citet{stawarz2019}, $\beta_e \approx 0.5$ and $\tau/Z\approx 10$; choosing $\nu=1$ and using the $a\gg\rho_s$ expression gives the critical length for ion-coupled reconnection $L/\rho_s\approx 10$ (this is also effectively the critical aspect ratio of such ion-coupled reconnecting sheets, since it is the point at which $a\approx\rho_s$). In other words, the fact that ion-scale sheets and ion jets are not often observed in the magnetosheath means that the aspect ratios of the sheets formed (presumably by the turbulent dynamics) are not extremely high. 

As $a\to\de$, current sheets of any appreciable aspect ratio greater than $1$ are disrupted -- so in practice, as one might have expected, the steady-state Sweet-Parker-like configuration, with width $d_e$ (which we derive in Appendix~\ref{sec:sp}) is not accessible. Reconnecting sheets for $a\ll\rho_s$ should have the aspect ratio given by the second expression in Eq.~(\ref{eq:lacrit}), and so we can estimate the aspect ratio of the sheets observed by \citet{phan2018}, with thicknesses $a\approx 4d_e$, to be no more than $\approx 4$, using the $n=1$ expression. %For these sheets, it is also possible to get an estimate of the disruption time $\tau_D \sim \gamma_{\rm max}^{-1}$ using the expression in Table \ref{tab:scalings}, along with $a\approx 4\mathrm{km}$ and $\vay \approx 25\mathrm{km/s}$, taken from \citet{phan2018}: $\tau_D \approx 0.04\mathrm{s}$.

A simple estimate of the disruption timescale of the current sheets observed by \citet{phan2018} is possible using Eq.~(\ref{eq:tstar}). Given that $aL$ is constant during collapse, we can estimate $a_0L_0 = aL \approx 64d_e^2$. Using $d_e \approx 1\mathrm{km}$, $\vay \approx 25\mathrm{kms^{-1}}$, and $\rho_i \approx 100\mathrm{km}$, $t_* \approx 0.06s$, which suggests that these electron-only reconnection events are individually rather short-lived.

\subsection{Outflow velocity}\label{sec:outflow}
One can obtain an estimate of the electron outflow velocity in the reconnecting sheets by a simple scaling argument. Our configuration is a current sheet of thickness $a$ and length $L\gg a$, and a known upstream field $B_{\rm in}$. The electrons flow into the sheet at $u_{\rm in}$, and out at $u_{\rm out}$ (which are related to $\phi$ via Eq.~(\ref{eq:frozen})). We will assume that $g_e = 0$ for simplicity; although this is not realistic, on the level of the scaling arguments presented here it is irrelevant.

First, because upon inspecting Eq.~(\ref{eq:frozen}), $\nablap \cdot \mathbf{u}_{\rm eff}=0$, 
\beq
L u_{\rm in} \sim a u_{\rm out}.\label{eq:divfree}
\eeq
Second, because $L\gg a$, the upstream energy is dominated by the magnetic field, while the energy downstream is dominated by the ion flow and the density perturbations \citep{zocco2011}. Balancing these,
\beq
\frac{B_{\rm in}^2}{8\pi} \sim \left[1+\zot\left(1-\hat\Gamma_0\right)\right]\zot\left(1-\hat\Gamma_0\right)\frac{e^2n_{0e}}{2\toe}\phi_{out}^2.
\eeq
In the limit $\rho_i\ll a$, $\zot(1-\hat\Gamma_0)\approx (\rho_s/a)^2$, and using Eq.~(\ref{eq:frozen}), this just results (reassuringly) in Alfv\'enic outflows, $u_{\rm out} \sim \vay$, where $\vay = B_{\rm in}/\sqrt{4\pi n_{0i}m_i}$. In the case where $\rho_i \gg a$, using $\hat\Gamma_0\approx 0$, we find
\beq
u_{\rm out} \sim \sqrt{\frac{1}{2}\left(1+\zot\right)}\vay \frac{\rho_i}{a},\label{eq:uout}
\eeq
which is just the in-plane kinetic-Alfv\'en-wave speed. So, rather different from the MHD case, the outflow velocity depends on the thickness of the sheet. Written in terms of the electron Alfv\'en speed, this is
\beq
u_{\rm out} \sim \sqrt{\frac{\beta_i}{2}\left(1+\zot\right)}\vaey \frac{d_e}{a}.\label{eq:ueout}
\eeq
For the \citet{phan2018} sheets, our model predicts (with $\beta_i\approx5$, cf. \citealt{stawarz2019}) $u_{\rm out}/\vaey\approx 0.4$, while the observed outflow speed in \citet{phan2018} is $u_{\rm out}/\vaey \approx0.25-0.45$; given the idealizations involved (not to mention the fact that we are pushing our equations rather beyond their low-$\beta$ regime of validity), this is reasonable agreement. More generally, we plot the dependence of the thickness at disruption, $a_{\rm crit}$, and the electron outflow velocity $u_{\rm out, crit}$, as functions of $(L/a)_{\rm crit}$ in Figure~\ref{fig:disrupt}.

Finally, it is worth asking what the effect of relaxing the low-$\beta$ ordering of the KREHM equations would be on the tearing mode and on the critical aspect ratio we obtain here. We have shown in Section \ref{sec:ermhdcollapse} that $\beta$ does not make a difference to the collapse process. \citet{numata2015} have shown that, if $\rho_i/a$ and $\rho_s/a$ are kept constant, the gyrokinetic tearing mode growth rate basically agrees with KREHM, i.e. proportional to $\beta_e^{-1/2}$. Thus, we believe that using the KREHM tearing mode scalings is physically (if not quite mathematically) reasonable up to $\beta_e\sim1$.

\section{Conclusions}
We have developed a theoretical model for low-$\beta$ current sheet formation and disruption by reconnection, and predict that sheets that are not sufficiently anisotropic will collapse to scales much smaller than the ion kinetic scales ($\rho_s$ and $\rho_i$), only reconnecting when their thickness is a few times the electron inertial scale $d_e$. For such sheets, we predict electron jets with velocities an appreciable fraction of the electron Alfv\'en velocity (i.e. much larger than the ion Alfv\'en velocity), and that there will be no ion-scale region and very slow or no ion outflows. As the aspect ratio of the sheets is varied, there is a gradual transition between the usual ``ion-coupled" regime and this new regime of ``electron-only" reconnection, with sheets reconnecting at thickness $a \approx \rho_s$ when their aspect ratio $L/a \approx 10$. We find that in the magnetosheath, this means that the aspect ratio of reconnecting sheets, observed to be of thickness $a \approx 4 d_e$, is no more than a factor of $\approx 4$. Our results may thus help to explain recent observations of electron-only reconnection in the magnetosheath \citep{phan2018}. 

Our picture of reconnection onset is based on two phenomena. First, ideal current sheets tend to collapse \citep{syrovatskii1981,chapman1963,biskampbook}, and we show in Section \ref{sec:collapse} that this collapse does not have to stop at the ion scale. Second, this ideal collapse process breaks down when the tearing mode growth rate becomes larger than the collapse (or sheet formation) rate, an application of the ideas of \citep{uzdensky2016} to a collisionless plasma. This implies a critical aspect ratio at which reconnection onset occurs, which we derive in Section \ref{sec:disrupt}. Importantly, the sheet thickness at disruption can be above or below the ion scales, and so we can predict the physical conditions in which we expect ion-coupled reconnection to give way to electron-only reconnection.

Our analysis is based on the KREHM equations, which are derived from gyrokinetics for $\beta_e \sim \beta_i\sim Zm_e/m_i\ll1$. This low-$\beta$ ordering is not usually formally valid in the magnetosheath. However, we can show that, in fact, the collapse process does not depend on the low $\beta$, and KREHM does a surpisingly good job of predicting the full gyrokinetic tearing rates even at moderate $\beta_e$. This perhaps explains why our estimates of electron outflow velocities agree reasonably well with the MMS observations. More generally, we believe that KREHM, while not always formally valid, is a useful minimal physical model of collisionless reconnection, which allows us to make theoretical progress in understanding electron-only reconnection and the transition from the ion-coupled to electron-only regimes.

\citet{boldyrevloureiro2019} use a different approach to the one used in this work: instead of studying a nonlinear collapsing solution, they show that if helicity in a structure is held constant, the nonlinear interaction is minimised by making the structure as sheet-like as possible: thus, sheet-like structures last longer. They then argue that this means that the turbulence favours the creation of highly sheet-like structures, but that this is limited by the tearing instability, which then sets a particular critical aspect ratio (as does our analysis). Despite the overall similarity of the results, there are some important differences in the two pictures: in ours, the nonlinear interaction actively produces increasingly sheet-like structures, while in \citet{boldyrevloureiro2019}, its role is to rapidly remove non-sheet-like structures. Our method has the advantage that it also works for sheets thicker than the ion scales, and we can therefore predict when one expects to see electron-only versus ion-coupled reconnection, depending on the characteristic aspect ratios of sheets as a function of their thickness.

In turbulence models that include the phenomenon of dynamic alignment \citep{boldyrev,chandran14,ms16}, the turbulent structures at scale $\lambda$ are sheet-like in the perpendicular plane, with aspect ratios $\propto (\lambda/L_\perp)^{-1/4}$, where $L_\perp$ is the outer scale at which energy is injected into the system. Thus, for electron-only reconnection to be observed in turbulence, energy must be injected relatively close to the ion scales: in the magnetosheath, this injection could potentially come in the form of Alfv\'en vortices \citep{Alexandrova2008}, perhaps driven by density gradients \citep{sundkvist2008}. If this picture is correct, electron-only reconnection may be relevant in other astrophysical environments with large density inhomogeneities, for example close to shocks. The appearance of electron-only reconnection in turbulent systems where the driving scale is relatively close to the ion scales has been observed rather clearly in the (2D) hybrid turbulence simulations of \citet{califano2018}, who observed electron-only reconnection when $d_i/L_\perp\approx 0.6$, but ion-coupled reconnection when $d_i/L_\perp \approx 0.3$. Our results therefore provide a theoretical context for these numerical results. 

Another low-$\beta_e$ environment of interest is the inner heliosphere and solar corona, currently being explored by Parker Solar Probe \citep{spp2016}. Our results show that whether ion-coupled or electron-only reconnection, or a mixture of the two, occur in this turbulent setting depends crucially on the length of the inertial range between the injection scale and the ion scales.

However, a detailed picture of electron-only reconnection onset in turbulence in the range of scales between $\rho_s$ and $d_e$ requires a model of how the aspect ratios of typical turbulent structures evolve in this range in the absence of reconnection (or, indeed, after ion-coupled reconnection decreases the typical aspect ratio, see e.g. \citealt{msc_cless}): this will be the focus of future work.

\section*{Acknowledgements} We thank A.~A.~Schekochihin, T.~D.~Phan, R.~Meyrand, D.~Gro{\v s}elj, and P.~Sharma~Pyakurel for useful conversations, and the anonymous reviewers for their helpful comments. This work was supported by NASA grant NNX16AP95G.

\appendix

\section{Steady-state reconnecting sheet}\label{sec:sp}
Let us perform a scaling analysis of Eqs.~(\ref{eq:mom}-\ref{eq:ge}), similarly to \citet{parker1957} and \citet{sweet1958}, and find a steady-state reconnecting configuration. The existence of this hypothetical configuration is quite reassuring, but as we have shown in Section \ref{sec:disrupt}, this steady-state configuration is not realizable, because it is violently unstable. This is reminiscent of the results of \citet{nuno}, \citet{pucci2014} and \citet{uzdensky2016} on the resistive plasmoid instability. We will work in 2D, i.e. $\partial/\partial z = 0$, and for simplicity set $\gel=0$. 
%We will additionally assume that, in terms of scalings, 
%\beq
%\phi \sim \left[1+\zot\left(1-\hat\Gamma_0\right)\right]\phi,\label{eq:approx}
%\eeq
%which is somewhat reasonable, since in the limit $k_\perp\rho_i\ll1$, $\left(1-\hat\Gamma_0\right)\ll1$, while in the limit $k_\perp\rho_i\gg1$, $(1-\hat\Gamma_0)\lesssim1$ so that the two sides of Eq.~(\ref{eq:approx}) only differ by a factor of order unity (i.e., we assume $Z/\tau\sim 1$). Given the other drastic simplifications in the following analysis, this is relatively reasonable.
Our configuration will the same as in Section \ref{sec:outflow}, namely a current sheet of thickness $a$ and a length $L\gg a$, with upstream field $B_{\rm in}$, and electrons flowing into the sheet at $u_{\rm in}$, and out at $u_{\rm out}$. The analysis of Eqs.~(\ref{eq:divfree}--\ref{eq:ueout}) to determine the outflow velocity in terms of $a$ is therefore still valid.

We additionally require a steady state, $\partial/\partial t = 0$. To achieve this, we balance the nonlinear terms in Eq.~(\ref{eq:apar}),
\beq
\left\{ \apar, \left[1+\zot\left(1-\hat\Gamma_0\right)\right]\phi \right\}\sim\left\{\phi,\de^2\nablap^2\apar\right\}.\label{eq:steady}
\eeq
This gives
\begin{align}
a\sim \left\{
\begin{array}{cr}
\de, &\de \gg \rho_i, \\
\frac{\de}{\sqrt{1+Z/\tau}},  &\de \ll \rho_i,\\
\end{array}
\right.\label{eq:aeqde}
\end{align}
using the small- and large-argument expansions of $\hat\Gamma_0$. %The equivalent balance for the (semi-)collisional case with field lines being broken by a resistive term $\eta \nablap^2\apar$ in Eq.~(\ref{eq:apar}) results in $u_{\rm in} \sim \eta/a$, which in combination with Eq.~(\ref{eq:divfree}) means that resistive Sweet-Parker sheets have a characteristic aspect ratio
%\begin{align}
%\frac{u_{\rm in}}{u_{\rm out}}\sim\frac{a}{L}\sim\left\{\begin{array}{lr}
%\sqrt{\frac{\eta}{L\vay}}\sim S_L^{-1/2}, & a\gg\rho_i, \\
%\frac{\eta}{\rho_i\vay}\sim S_{\rho_i}^{-1}, & a \ll\rho_i, \\
%\end{array}\right.
%\end{align}
%ignoring factors of order unity involving $1+Z/\tau$ for simplicity. Thin sheets described by the second equation are only possible if their lengths also satisfy $L/\rho_i \ll S_L^{1/2}$, $S_L$ being the Lundquist number $S_L = L\vay/\eta$\footnote{Interestingly, this expression is independent of $a$ and $L$, meaning that semicollisional Sweet-Parker sheets have a constant aspect ratio. The dimensionless reconnection rate as usually defined is $u_{\rm in}/\vay\sim \rho_i/L$, independent of $\eta$.}.
Thus, these steady-state sheets have a constant thickness $a\sim\de$, and an arbitrary length $L$, i.e.
\beq
\frac{u_{\rm in}}{u_{\rm out}} \sim \frac{\de}{L}.\label{eq:sp_de}
\eeq
The dimensionless reconnection rate as usually defined is
\beq
R \sim \frac{u_{\rm in}}{\vay} \sim\left\{\begin{array}{lr}
{\de}/{L}, & \de\gg\rho_i,\\
{\rho_i}/{L}, &\de\ll\rho_i.\\
\end{array}
\right.
\eeq
We have managed to derive a Sweet-Parker-like scenario for collisionless reconnection. The result for $\de\gg\rho_i$ has been previously discovered by \citet{wesson1990}, while \citet{bulanov1992} and \citet{avinash1998} also derived the scaling (\ref{eq:sp_de}) for electron MHD (i.e., in the whistler frequency range).

Note that in the case $d_e \ll \rho_i$, the outflow velocity is
\beq
u_{\mathrm{out}} \sim \sqrt{\frac{\beta_i}{2}\left(1+\zot\right)}\vaey,
\eeq
of order the electron Alfv\'en velocity $\vaey$. In Section \ref{sec:disrupt}, we show that this steady-state configuration is not realizable, because reconnection onset occurs at a somewhat larger thickness.

\section{Tearing mode in a sheet with thickness below the ion scales}\label{sec:tear}
Let us analyze the tearing mode of Eqs.~(\ref{eq:mom}-\ref{eq:ge}) -- we will study the 2D case, $\partial/\partial z = 0$, and use as our equilibrium
\beq
\phi_0=0, \quad g_{e0} = 0, \quad A_{\parallel0} = A_{\parallel0}(x),
\eeq
where $A_{\parallel0}(x)$ is controlled by a dimensionless function $f(x)$, varying on the equilibrium scale-length $a$,
\beq
B_{y0} = -\frac{dA_{\parallel0}}{dx} = b_a f(x).
\eeq
Due to the symmetry of this equilibrium, our perturbations take the form
\begin{align}
\left.\begin{array}{l}
\phi_1 \\
A_{\parallel1}\\
g_{e1}\\
\end{array}
\right\}=\left\{
\begin{array}{l}
\phi(x) \\
\apar(x)\\
\gel(x)
\end{array}
\right\} e^{ik_y+\gamma t}
\end{align}
We will discuss two different regimes,
\begin{align}
\text{(i)}&\quad a \gg \rho_s \gg d_e, \\
\text{(ii)}&\quad \rho_s\gg a \gg \de,
\end{align}
where (i) has already been treated by ZS11 (and references therein), while (ii) represents the tearing of sub-ion-scale sheets we are interested in the present work.

The linearized equations may be written\footnote{Neglecting $k_y^2\de^2$ and $\de^2f''/f$ compared to unity.}
\begin{align}
\frac{2}{\de^2}\frac{\delta}{af}\zot \left(1-\hat\Gamma_0\right)\tilde{\phi} = \left[\partial_x^2 -k_y^2 - \frac{f''}{f}\right]\apar, \label{eq:philin}\\
\frac{\delta}{af}\left(1-\de^2\partial_x^2\right)\apar = \left[1+G\zot\left(1-\hat\Gamma_0\right)\right]\tilde{\phi}\label{eq:aparlin}
\end{align}
where
\beq
\delta = \frac{\gamma a B_0}{k_y b_a \vthe},\quad \tilde\phi=-\frac{c}{\vthe}i\phi,
\eeq
and the function $G$ which results from solving the linearized $\gel$ equation \citep{zocco2011} is
\beq
G\left(\frac{a|f|}{\delta}\right)= -2 \left[\frac{\delta^2}{a^2f^2} + \frac{1}{Z'({i\delta}/{a|f|})}\right],
\eeq
$Z$ here being the plasma dispersion function. This has limiting values $G(0) = 3$, $G(\infty)=1$. We will make further progress by, as usual, solving separately in an ``outer region'' where $x\sim a$ and an ``inner region" where $x\ll a$.
\subsection{Outer region}
Here, $x\sim a$ and $G\approx 1$, and $\pxpx \sim f''/f \sim 1/a$. We may therefore neglect $\de^2\pxpx\ll1$ in Eq.~(\ref{eq:aparlin}). The structure of the equation depends on $a/\rho_s$. In case (i), $a\gg \rho_s$, one can expand $\hat\Gamma_0$ for small argument and show that the LHS of Eq.~(\ref{eq:philin}) is smaller than the RHS by $\gamma/\omega_{Ay}$, where $\omega_{Ay}=k_y b_a f/\sqrt{4\pi{n_{0i}}{m_i}}$ is the in-plane Alfv\'en frequency; assuming this is small (which may be checked afterwards), one just obtains the same outer region equation as in the MHD tearing mode:
\beq
\left[\pxpx - k_y^2 -\frac{f''}{f}\right]\apar=0.\label{eq:mhdouter}
\eeq
In case (ii), however, where $a\ll\rho_s$, $\hat\Gamma_0\approx 0$ and Eq.~(\ref{eq:aparlin}) becomes
\beq
\frac{\delta}{af}\apar = \left[1+\zot\right]\tilde{\phi}.
\eeq
Substituting this into Eq.~(\ref{eq:philin}) (again taking $\hat\Gamma_0\approx 0$),
\beq
\lambda^2\frac{2}{f^2}\frac{1}{\frac{\tau}{Z}+1} \apar = a^2\left( \partial_x^2 - k_y^2 - \frac{f''}{f}\right)\apar,\label{eq:outerii}
\eeq
where $\lambda^2 = \delta^2/\de^2$. The LHS may only be neglected in the outer region (where $f\sim1$) if $\lambda^2\ll 1$. In this case, one can see that, because $\delta B_y =-\partial_x \apar$ must change sign at $x=0$,
\beq
\apar^{\rm outer} = \apar(0)\left(1+\frac{1}{2}\Delta'|x|\right) \text{ for } x \ll a.
\eeq
We will determine when this is valid later. For $k_ya \ll 1$, $\Delta'a \propto (k_y a)^{-n}$ with $1\leq n\leq 2$\footnote{$n=1$ corresponds to, for example, a ``Harris"-type equilibrium $f=\tanh(x/a)$, while an example of an $n=2$ equilibrium is $f=\sin(x/a)$.}.

\subsection{Inner region}
Here, $x\ll a$, $\pxpx\gg k_y^2,f''/f$, and $f \approx x/a$. Our analysis will at first follow ZS11 exactly. The equations become
\begin{align}
\frac{2}{\de^2}\frac{\delta}{x}\zot\left(1-\hat\Gamma_0\right)\tilde\phi&=\pxpx\apar,\\
\frac{\delta}{x}\left(1-\de^2\pxpx\right)\apar &= \left[1+G\zot\left(1-\hat\Gamma_0\right)\right]\tilde\phi,\label{eq:aparinner1}
\end{align}
where $\hat\Gamma_0=\hat\Gamma_0(\pxpx)$. Let us rescale the coordinate to $\xi = x/\delta_{\rm in}$, where $\delta_{\rm in}$ will be chosen later. Replacing the term involving $\hat\Gamma_0$ in Eq.~(\ref{eq:aparinner1}),
\begin{align}
\frac{2\delta\delta_{\rm in}}{\de^2}\zot\left(1-\hat\Gamma_0\right)\tilde\phi&=\xi\apar'',\label{eq:phiinner2}\\
\frac{\apar}{\xi} - \frac{\din}{\delta}\tilde\phi &= \frac{\de^2}{\din^2}\left(1+\frac{\din^2}{2\delta^2}G\xi^2\right)\frac{\apar''}{\xi},\label{eq:aparinner2}
\end{align}
where primes denote derivative with respect to $\xi$. If one wants to include the case where $\din\sim\rho_s$, the operator $\hat\Gamma_0$ is difficult to deal with analytically: we will follow ZS11 and use the Pad\'e approximant
\beq
\zot\left(1-\hat\Gamma_0\right) \approx -\frac{\rho_s^2\pxpx}{1-(1/2)\rho_i^2\pxpx}.\label{eq:pade}
\eeq
Let us make the helpful substitution $\chi=\xi\apar' - \apar=\xi^2(\apar/\xi)'$, so that $\chi'=\xi\apar''$. Differentiating Eq.~(\ref{eq:aparinner2}), using the approximant (\ref{eq:pade}) in Eq.~(\ref{eq:phiinner2}) and integrating once, the equations in terms of $\chi$ are
\begin{align}
-\frac{2\delta\rho_s^2}{\de^2\din}\tilde\phi'&= \chi-\chi_0-\frac{\rho_i^2}{2\din^2}\chi'', \\
\frac{d}{d\xi}\left(\frac{1}{\xi^2}+\frac{\din^2}{2\delta^2}G\right)\chi'&=\frac{\din^2}{\de^2}\frac{\chi}{\xi^2}-\frac{\din^3}{\de^2\delta}\tilde\phi'.
\end{align}
Combining these two equations, we get
\beq
\xi^2\frac{d}{d\xi}\left[\frac{1}{\xi^2}+\frac{\din^2}{2\delta^2}\left(G+\toz\right)\right]\chi'=\frac{\din^2}{\delta^2}\left[\lambda^2\chi + \frac{1}{2}\xi^2\frac{\din^2}{\rho_s^2}\left(\chi-\chi_0\right)\right].\label{eq:innerchi}
\eeq
Noting that $\xi\sim1$ in the inner region, we can compare the size of the two terms on the RHS. 

In case (i), $a\gg\rho_s$, we must allow a finite $\din/\rho_s$ and must keep both of them. It therefore makes sense to define $\din = (\sqrt{2}\rho_s\delta)^{1/2}$, $\alpha = \din/\sqrt{2}\delta$, $\tilde\lambda^2 = \lambda^2\alpha^2$, and we obtain the same equations as in ZS11,
\beq
\xi^2\frac{d}{d\xi}\left[\frac{1}{\xi^2}+\frac{\alpha^2}{2}\left(G+\toz\right)\right]\chi'=\tilde{\lambda}^2\chi + \xi^2\left(\chi-\chi_0\right).\label{eq:zs11inner}
\eeq
It is then possible to solve this analytically in terms of a nested ``ion inner region" and ``electron inner region", matched to each other and to the outer solution, via the matching condition
\beq
\frac{1}{\chi_0}\int_0^\infty \frac{\chi'}{\xi} d\xi = -\frac{1}{2}\Delta'\din.\label{eq:matching}
\eeq
One can also (again, following ZS11) guess the scalings of the solution on dimensional grounds. The width of $\chi'/\xi$ in the integral above is $x\sim\delta$, or $\xi \sim 1/\alpha$. From Eq.~(\ref{eq:zs11inner}), $\chi'/\xi\sim\tilde\lambda^2\chi$, whence $\tilde\lambda^2 \sim \Delta'\din\alpha$. For $\Delta'\din\gg1$, the current cannot depend on $\Delta'$, and we cannot use (\ref{eq:matching}); instead $\chi'/\xi \sim \chi$ (the current is limited by $\din$, $\apar'' \sim \apar/\din^2$), and so $\tilde\lambda^2 \sim \alpha$. The resulting growth rates $\gamma$, inner widths $\din$, and electron region widths $\delta$ are shown in table \ref{tab:scalings}. The maximum growth rate, obtained by inserting $\Delta'\din \sim 1$ into the small-$\Delta'$ expression (or alternatively by balancing the small- and large-$\Delta'$ growth rates), is also shown.

One might also be interested in the dependence of the growth rates on $\tau/Z$. It is possible to solve this equation analytically (see ZS11), showing that in fact, the growth rates are proportional to
\beq
H(\tau/Z) = \sqrt{1+\tau/Z}.\label{eq:H}
\eeq
This factor is also shown in Table~\ref{tab:scalings}.
%\begin{table}\label{tab:scalings}
%\begin{center}
%\begin{tabular}{lcccccc}
% $a/\rho_s$&$\Delta'\din$&$\gamma/k_y\vay$& $\delta$ & $\din$&$\gamma_{\rm max}a/\vay$&$k_{\rm max}a$  \\
% \hline
%\multirow{2}{*}{$\gg 1$}&$\ll1$&$\Delta' {\de\rho_s}/{a}$& $\de^2\Delta'$&$\de\rho_s^{1/2}\Delta'^{1/2}$& \multirow{2}{*}{$\frac{\de^{1/3+2/3n}\rho_s^{2/3+1/3n}}{a^{1+1/n}}$}&\multirow{2}{*}{$\left(\frac{\de^{2/3}\rho_s^{1/3}}{a}\right)^{1/n}$}   \\
%&$\gg1$&${\rho_s^{2/3}\de^{1/3}}/{a}$&${\de^{4/3}}{\rho_s^{-1/3}}$&$\rho_s^{1/3}\de^{2/3}$&   \\
%\multirow{2}{*}{$\ll 1$}&  $\ll1$&$\Delta' {\de\rho_s}/{a}$& $\de^2\Delta'$&$\delta$&\multirow{2}{*}{$\frac{\rho_s}{a}\left(\frac{\de}{a}\right)^{1/n}$}&\multirow{2}{*}{$\left(\frac{\de}{a}\right)^{1/n}$}    \\
%&($\gg1$&${\rho_s}/{a}$&$\de$&$\delta$)&    \\
%
%\end{tabular}
%\caption{Scalings for the collisionless tearing mode in various limits. The final row, $a/\rho_s\ll1$ and $\Delta'\din \gg 1$, is probably inaccurate since we do not have an analytic solution.}
%\end{center}
%\end{table}

In case (ii), $a\ll \rho_s$, there is no space for the ion inner region to form, and we must instead match the solution in the electron inner region directly onto the outer solution. Going back to Eq.~(\ref{eq:innerchi}), if $\din \ll \rho_s$ (as it must be since $\rho_s\gg a \gg \din$), we can neglect the second term on the RHS, and set $\din = \delta$, obtaining
\beq
\xi^2\frac{d}{d\xi}\left[\frac{1}{\xi^2}+\frac{1}{2}\left(G+\toz\right)\right]\chi'=\lambda^2\chi.\label{eq:innerchismall}
\eeq
We can again guess the scalings on dimensional grounds, similarly to in case (i). For $\Delta'\delta\ll1$, $\lambda^2 \sim \Delta'\delta$. Therefore, we may solve Eq.~(\ref{eq:innerchismall}) perturbatively in powers of $\lambda^2\ll1$, to first order, i.e. $\chi=\chi_0+\chi_1$, with the boundary condition that the current $\chi'/\xi$ is even as $\xi\to0$. Setting the RHS to zero, the $0$th order solution $\chi_0$ is just a constant. This means our solution is equivalent to taking a ``constant-$\psi$" approximation \citep{fkr}. At the next order, using $\chi_0$ on the RHS and the boundary condition, we obtain
\beq
\frac{\chi_1'}{\xi} = -\frac{\lambda^2\chi_0}{1+\xi^2/2(G+\tau/Z)},
\eeq
and matching this to the outer solution using Eq.~(\ref{eq:matching}), 
\beq
\lambda^2 = \frac{\Delta'\delta}{2I}
\eeq
where
\beq
I=\int_0^\infty\frac{d\xi}{1+\xi^2/2(G+\tau/Z)}.
\eeq
This is precisely the same growth rate, including the prefactor, as for the small-$\Delta'$ case with $a\gg\rho_s$ (cf. \citet{zocco2011}, Equation B71). For $\tau/Z\gg1$, we can neglect $G$ and $I\sim \sqrt{Z/\tau}$, while for $\tau/Z\ll1$, $I\sim 1$. We include this effect in Table \ref{tab:scalings} by amending the scalings by the factor $H(\tau/Z)$ defined in Eq.~(\ref{eq:H}).

For $\Delta'\delta \gg 1$, on dimensional grounds, $\lambda^2\sim 1$: unfortunately, this invalidates the neglect of the LHS in Eq.~(\ref{eq:outerii}), meaning that our matching condition (\ref{eq:matching}) is no longer valid. We have not yet obtained an analytic solution to the equations in this case. However, since on physical grounds the growth rate must decrease as $k_y\to 0$, it is reasonable to assume that the maximum growth rate of the tearing mode may be estimated by setting $\Delta'\delta\sim 1$ in the scalings obtained for the $\Delta'\delta \ll 1$ limit. The scalings derived here for the tearing mode with $a\ll\rho_s$ are also shown in Table \ref{tab:scalings}. 

\citet{boldyrevloureiro2019} have derived a somewhat similar tearing mode for the ``inertial kinetic-Alfv\'en" regime \citep{chen2017}. The difference between their equations and the KREHM equations \citep{zocco2011} is that theirs are fluid (i.e. they do not contain heating), and, secondly, they order $\beta_i\sim 1$, which introduces an additional term in our Eq.~(\ref{eq:phiinner2}) (their Eq.~18) proportional to $\beta_i\phi''$ ($b_z''$ in their variables). This means they can neglect the term on the RHS in our equation, because in the inner region the gradients are large, and then makes their equations structurally identical to the MHD tearing equations. They therefore obtain different maximum growth rates $\gamma_{\rm max} \sim (d_e/a)^{1+1/n}\vaey/a$ that are a multiplied by a factor $\beta_i^{1/2}$ compared to ours.

\bibliographystyle{jpp}
\bibliography{mainbib2}
\end{document}